\documentclass[aps,pre,reprint,groupedaddress,showpacs]{revtex4-1}
\usepackage{graphicx}

\newcommand{\Fig}[1]{Fig.~(\ref{#1})}
\newcommand{\Eq}[1]{Eq.~(\ref{#1})}
\newcommand{\eq}[1]{Eq.~(\ref{#1})}
\newcommand{\eqr}[1]{(\ref{#1})}
\renewcommand{\vec}[1]{\mathbf{#1}}

\newcommand{\Gam}{\frac{\gamma}{2}}

\newcommand{\Order}[1]{\mathcal{O}(#1)}

\begin{document}


\title{Radial propagation in population dynamics with density-dependent diffusion}


\author{Waipot Ngamsaad}
\email{waipot.ng@up.ac.th}
\affiliation{Division of Physics, School of Science, University of Phayao, Mueang Phayao, Phayao 56000, Thailand}


\date{\today}

\begin{abstract}
The population dynamics that evolves in the radial symmetric geometry is investigated. The nonlinear reaction-diffusion model, which depends on population density, is employed as the governing equation for this system. The approximate analytical solution to this equation has been found. It shows that the population density evolves from initial state and propagates as the traveling wave-like for the large time scale. One can be mentioned that, if the distance is insufficient large, the curvature has ineluctable influence on density profile and front speed. In comparison, the analytical solution is in agreement with the numerical solution.
\end{abstract}


\maketitle

The growth and dispersal of species in populations undergo the density spreading as the traveling wave front \cite{Murray2002}. This phenomenon becomes an active research topic for many decades. In theoretical framework, the dynamics of population can be modeled as diffusion with reaction processes. The paradigmatic model is known as the Fisher equation \cite{Fisher1937}, which has been originated as a model for the population genetics \cite{Aronson1975, Aronson1978, Newman1980, Newman1983, Murray2002}. The solution to this equation has demonstrated the propagating as the traveling wave front in population dynamics \cite{Fisher1937, Murray2002, Volpert2009}. This equation and its variant have also appeared in various systems, including chemical dynamics \cite{Murray2002}, nerve pulse propagation \cite{Aronson1975}, flow in porous media \cite{Aronson1986}, combustion theory \cite{Aronson1975, Newman1980, Newman1983}, wound healing \cite{Sherratt1990, Maini2004}, tissue engineering \cite{Sengers2007, Simpson2013} and bacterial pattern formation \cite{Kawasaki1997, Ben-Jacob2000, Murray2002}. 

In the original Fisher model \cite{Fisher1937}, the evolution of population density $u(\vec{r},t)$, at spatial position $\vec{r}$ and time $t$, is governed by the simplest nonlinear reaction-diffusion equation \cite{Fisher1937, Murray2002}. The reaction term is modeled as logistic law and it describes the growth of population with limited supply. The movement of individual is modeled as random walk \cite{Skellam1951}, where the diffusion coefficient is constant. However, the motion of the biological population is not purely random but they move with sense. To remedy this issue, the directed motion model, in which individuals tend to move in the direction of decreasing populations as fast as increasing density, has been  proposed \cite{Gurney1975, Gurney1976}. The diffusion coefficient in this model depends on the population density \cite{Gurney1975, Gurney1976, Gurtin1977, Newman1980, Newman1983}. Later, a general form of the logistic law has been found \cite{Newman1983}. With these modifications, the density-dependent reaction-diffusion equation or the extended Fisher model has been presented \cite{Newman1980, Newman1983, Witelski1995, Murray2002, Ngamsaad2012},
\begin{equation}\label{eq:gen_fisher_dimless_full}
\frac{\partial u}{\partial t} = \nabla\cdot \left[D\left(\frac{u}{u_M}\right)^p\nabla u\right] + \alpha u\left[1 - \left(\frac{u}{u_M}\right)^p\right] ,
\end{equation} 
where $p>0$, $D$ is diffusion constant, $\alpha$ is rate constant and $u_M$ is maximum population density. 

The solution of \eq{eq:gen_fisher_dimless_full} in one dimension (1D) is known as sharp traveling wave, propagating with constant front speed \cite{Newman1980, Newman1983, Rosenau2002, Murray2002}. In our previous work, a general form of solution to \eq{eq:gen_fisher_dimless_full} in  one-dimensional space have been found \cite{Ngamsaad2012}. This solution shows that the population density evolves, from a specific initial condition, as a self-similar pattern that converges to the traveling wave at large time scale \cite{Ngamsaad2012}. Although the solution of \eq{eq:gen_fisher_dimless_full} in one-dimensional space has been known \cite{Newman1980, Newman1983, Rosenau2002, Murray2002, Ngamsaad2012}, its behavior in higher dimensions has not well understood. Typically, the population dynamics takes place in two dimensions (2D), sometimes in three dimensions (3D). Therefore, the solution of \Eq{eq:gen_fisher_dimless_full} in dimension higher than one could provide better insight into the dynamics of population. 

In this work, we study the population dynamics that evolves with radial symmetric geometry. In this form, the system is governed by the extended Fisher equation \eqr{eq:gen_fisher_dimless_full} in axisymmetric coordinate system. Before describing further, we change following quantities to be dimensionless: $u^\prime = u/u_M$, $t^\prime = \alpha t$ and $\vec{r}^\prime = \sqrt{(p+1)\alpha/D}\vec{r}$. Then, the radial symmetric extended Fisher equation in dimensionless form is given by 
\begin{equation}\label{eq:gen_fisher_full_rad1}
\frac{\partial u}{\partial t} = \frac{\partial^2 u^m}{\partial r^2} + \frac{\gamma}{r}\frac{\partial  u^m}{\partial r}  +u - u^m ,
\end{equation} 
where $m=p+1$, $r=|\vec r|$, $0 \leq r < \infty$, $\gamma = N-1$ and $N$ is dimension. Here, the prime symbols are dropped for convenience. \Eq{eq:gen_fisher_full_rad1} recovers dynamics in 1D when $\gamma=0$. Since the exact solution of \Eq{eq:gen_fisher_full_rad1} in 1D has been found \cite{Ngamsaad2012}, we focus on its solution in 2D, as well as in 3D. 

\Eq{eq:gen_fisher_full_rad1} does not support the traveling wave solution because the presence of the gradient term $(\gamma/r)\partial u^m/\partial r$ \cite{Murray2002}. It reduces to 1D problem at $r\to\infty$, which has the planar traveling wave as solution. Nevertheless, the behavior of this system at the distance that is not so large has been unclear. It has been mentioned that the effects of curvature can make the front propagation in reaction-diffusion system somewhat to be different from the planar case \cite{Volpert2009, Witelski2000}. Previously, \Eq{eq:gen_fisher_full_rad1} in cylindrical coordinate has been analyzed by the perturbation method \cite{Witelski1995}. However, the solution is shown in the large distance that yields the usual traveling wave. More recent, the Lie symmetry method has been employed to solve \Eq{eq:gen_fisher_full_rad1} for $m=2$ in cylindrical coordinate ($\gamma=1$) \cite{Bokhari2008}. Although the exact solution has been found, it is another class and does not reflect the density distribution of population. In this work, we adapt the technique similar to our previous studies \cite{Ngamsaad2012, Ngamsaad2012b} to solve for the solution of \eq{eq:gen_fisher_full_rad1}. We have found the approximate radial symmetric solution for \Eq{eq:gen_fisher_full_rad1} in the intermediate regime that the distance is not so large. The solution reveals the curvature effect on the spreading of the population, both of density profile and front speed, in this regime explicitly. To verify the analytical solution, we have solved \Eq{eq:gen_fisher_full_rad1} by a numerical method. The numerical result seems to agree with this approximate solution. It confirms that our approximate solution is plausible to describe the intermediate behavior of system. 

Before finding the solution in general case, we study the asymptotic behavior of \Eq{eq:gen_fisher_full_rad1}, as time goes to infinity, first. \Eq{eq:gen_fisher_full_rad1} can be viewed as a one-dimensional nonlinear convection-reaction-diffusion equation with the varying drift coefficient $\gamma/r$, similar to \cite{Uchiyama1985}. The gradient term $(\gamma/r)\partial u^m/\partial r$ in \Eq{eq:gen_fisher_full_rad1} is large in the vicinity of front position $R(t)$, otherwise it becomes small \cite{Volpert2009}. Thus, we change the gradient term to $(\gamma/R)\partial u^m/\partial r$. The drift coefficient becomes small as $R\to\infty$. In another hand, the nonlinear convection-reaction-diffusion equation with constant drift coefficient $\nu$ has been analyzed previously  \cite{Gilding2005, Mansour2010, Ngamsaad2012b}. It has found that the solution in this case, at large time, converges to the sharp traveling front at speed $c = \sqrt{1+\left(\nu/2\right)^2}-\nu/2$ \cite{Gilding2005, Mansour2010, Ngamsaad2012b}. For the small varying drift coefficient, we assume that the approximate front speed for \Eq{eq:gen_fisher_full_rad1} can be obtained by setting $\nu=\gamma/R$, 
\begin{equation}\label{eq:front_bias_velocity}
c = \sqrt{1+\left(\frac{\gamma}{2R}\right)^2}-\frac{\gamma}{2R} .
\end{equation}
The front speed \eqr{eq:front_bias_velocity} approaches to 1  as $R\to\infty$, which is equal to the constant front speed of planar wave \cite{Newman1980, Newman1983, Rosenau2002, Murray2002, Ngamsaad2012}. If the solution of \Eq{eq:gen_fisher_full_rad1} exists, it should result the front speed \eqr{eq:front_bias_velocity} as the asymptotic behavior.

We now perform the analysis to find the general form of density profile that propagates at the front speed of \Eq{eq:front_bias_velocity}. In approximation, we rewrite \Eq{eq:gen_fisher_full_rad1} in the form of
\begin{equation}\label{eq:gen_fisher_rad3}
\frac{\partial u}{\partial t} = \left(\frac{\partial}{\partial r} + \kappa^\ast \right)\left(\frac{\partial}{\partial r} - \kappa \right)u^m +u +\frac{\partial \kappa}{\partial r}u^m , 
\end{equation} 
where 
\begin{eqnarray}
\label{eq:v1}
\kappa(r) &=& \sqrt{1+\left(\frac{\gamma}{2r}\right)^2}-\frac{\gamma}{2r} , \\
\label{eq:v2}
\kappa^{\ast}(r) &=& \sqrt{1+\left(\frac{\gamma}{2r}\right)^2}+\frac{\gamma}{2r} .
\end{eqnarray}
We note that $\kappa^\ast-\kappa=\gamma/r$ and $\kappa^\ast \kappa=1$. The correction term in \Eq{eq:gen_fisher_rad3}, $\partial \kappa/\partial r = \left\lbrace 1-\left[1+\left(2r/\gamma\right)^2\right]^{-1/2}\right\rbrace\frac{\gamma}{2r^2}$, approaches to $\Order{1/r^2}$ for $r\gg\gamma/2$ and to $1/\gamma + \Order{r^2}$ for $r\ll\gamma/2$. Fortunately, this correction term well behaves because it decays from $1/\gamma$ to zero as $r\gg 0$. In addition, at $r \ll \gamma/2$, the correction term does not much affect the initial state while $u\ll 1$. Next, we introduce the transformation $d\eta = dr/\kappa$, which can be evaluated to
\begin{equation}\label{eq:r_curve}
\eta(r) = \kappa r + \Gam\ln\left( \kappa r \right) .
\end{equation}
With the transformation \eqr{eq:r_curve}, \Eq{eq:gen_fisher_rad3}, by dropping the correction term, becomes
\begin{equation}\label{eq:gen_fisher_rad_tran1}
\frac{\partial u}{\partial t} = \kappa^{-1}\left(\frac{\partial}{\partial \eta} + 1 \right) \kappa^{-1} \left(\frac{\partial}{\partial \eta} - \kappa^2 \right)u^m +u  .
\end{equation} 
For $r\gg\gamma/2$, we approximate that $\kappa \approx 1 + \Order{1/r}$. Applying this approximation to \Eq{eq:gen_fisher_rad_tran1}, we obtain
\begin{equation}\label{eq:gen_fisher_rad_tran2}
\frac{\partial u}{\partial t} \approx \frac{\partial^2 u^m}{\partial \eta^2} +u -u^m . 
\end{equation} 
\Eq{eq:gen_fisher_rad_tran2} is equivalent to \Eq{eq:gen_fisher_full_rad1} in 1D, but evolving with $\eta$ as the spatial coordinate.

By adapting the result from Ref. \cite{Ngamsaad2012}, we obtain the solution to \Eq{eq:gen_fisher_rad_tran2},
\begin{eqnarray}\label{eq:main_sol_exact}
\lefteqn{
u(r,t) = \frac{\rho e^{t}}{\left[ \rho^p\left(e^{pt} - 1\right) + 1 \right]^\frac{1}{p}} 
}\nonumber\\ 
&& \times \left\lbrace 1 -  \left[\frac{e^{p \left(\eta(r)-\eta_0\right)}}{ \rho^p\left(e^{pt} - 1\right) + 1 }\right]^{\frac{1}{p+1}} \right\rbrace ^\frac{1}{p} ,
\end{eqnarray}
where $\eta_0 = \eta(r_0)$, $r_0$ is initial front position and $\rho$ is initial density amplitude. By setting $\gamma=0$, \Eq{eq:main_sol_exact} recovers the solution in 1D \cite{Ngamsaad2012}. We note that $u(r,t)$ vanishes after front position for $r\geq R(t)$, which will be determined later.  As $r\to 0$, we have $\eta\to -\infty$. This causes the density profile at the origin approaches to $u(0,t) = \rho e^{t}/\left[ \rho^p\left(e^{pt} - 1\right) + 1 \right]^\frac{1}{p}$, which is actually the solution of
\begin{equation}\label{eq:gen_fisher_full_zero}
\frac{\partial u(0,t)}{\partial  t}   =   u(0,t) - u^m(0,t) .
\end{equation} 
This implies no diffusion at the origin.

At a sufficient large time that $e^{pt^\prime}\gg 1$ and consequently $\rho^p e^{pt^\prime} \gg 1$, we estimate the transition point 
\begin{equation}\label{eq:transition_t}
t^{\prime} \approx  -\ln \rho . 
\end{equation}
For $t \gg t^\prime$, the solution \eqr{eq:main_sol_exact} emerges a pattern form of the traveling wave-like 
\begin{equation}\label{eq:main_sol_exact_traveling_wave}
\widetilde u(r,t) = \left\lbrace 1 - \left[\rho^{-1}e^{\left( \eta(r) -t -\eta_0 \right)}\right]^{\frac{p}{p+1}} \right\rbrace ^\frac{1}{p} .
\end{equation}
For $r\gg\gamma/2$, we approximate that $\eta(r) \approx r + \Gam\ln r$. It is seen that the logarithmic term does not vanish even at large distance, unless $\gamma = 0$. Therefore, \Eq{eq:main_sol_exact_traveling_wave} contains ineluctable curvature term, which can be called the curved traveling wave-like.

The front position can be calculated from \Eq{eq:main_sol_exact_traveling_wave} by determining the first position $R(t)$ that density falls to zero or  $\widetilde u(R,t)=0$. After evaluating, we obtain equation for front position
\begin{equation}\label{eq:front_position}
\eta(R) - \eta_0 =  t - t^\prime .
\end{equation}
From \eq{eq:front_position}, we see that the front position does not simply linearly depend on time as in 1D case \cite{Ngamsaad2012}. By differentiating, respected to time, both sides of \Eq{eq:front_position}, we obtain $\frac{\partial \eta(R)}{\partial R}\frac{dR}{dt} = 1$. This allow us to calculate the front speed $c = \frac{dR}{dt}$, that is
\begin{equation}\label{eq:front_speed}
c   =  \left(\frac{\partial \eta(R)}{\partial R}\right)^{-1} = \sqrt{1+\left(\frac{\gamma}{2R}\right)^2}-\frac{\gamma}{2R} .
\end{equation}
The front speed \eqr{eq:front_speed} obtained from this analysis recovers \Eq{eq:front_bias_velocity} as expected. Once again, we have seen that the front speed is altered by the curvature as found in other similar systems \cite{Witelski2000, Volpert2009}. At sufficient large distance that $R \gg \gamma/2$, the front speed can be approximated as a constant $c \approx 1$. This is equal to the front speed in  1D case ($\gamma=0$) \cite{Newman1980, Newman1983, Rosenau2002, Murray2002, Ngamsaad2012}.  

If we define the following quantities: $\phi(r) = e^{(p+2) \left(\eta(r)-\eta_0\right)/(p+1)}$, $\tau(t)=\rho^p\left(e^{pt} - 1\right) + 1 $, and $u(r,t)=\rho e^{t} e^{\left(\eta(r)-\eta_0\right)/(p+1)} w(r,t)$, we can rewrite \Eq{eq:main_sol_exact} as the scaling function
\begin{equation}\label{eq:scaling_law}
w(\phi,\tau) = \frac{1}{\tau^\beta}F\left(\frac{\phi}{\tau^\beta}\right) ,
\end{equation}
where $\beta = \frac{p+2}{p(p+1)}$, $F(\xi)=\left[\xi^{-p/(p+2)}-1\right]^{1/p}$ and $\xi=\phi/\tau^{\beta}$. In the term of transformed density $w(\phi,\tau)$, as a function of transformed space $\phi$ and time $\tau$, evolving of the population density in the radial symmetric geometry still holds the self-similarity with the scaling law of \Eq{eq:scaling_law}. Moreover, this self-similar pattern converges to the traveling wave-like \eqr{eq:main_sol_exact_traveling_wave} as time becomes large. The connection between self-similar solution and traveling wave solution can be described as intermediate asymptotics of the system \cite{Barenblatt1972}.

To compare with the analytical solution, we employ the standard explicit finite difference scheme \cite{NumericalRecipes} to solve the radial symmetric extended Fisher equation \eqr{eq:gen_fisher_full_rad1} numerically. \Eq{eq:gen_fisher_full_zero} is imposed as the boundary condition at origin ($r=0$). The boundary condition at the edge of computational domain is free, as the front never reaches to this position. The initial density profile for the numerical calculation is set to the same value of the analytical one, $u(r,0)$. The initial front position is chosen such that $r_0\gg\gamma/2$, since the analytical solution is expected to be accurate at large distance.

\begin{figure}[h]
\centerline{\includegraphics[width=\columnwidth]{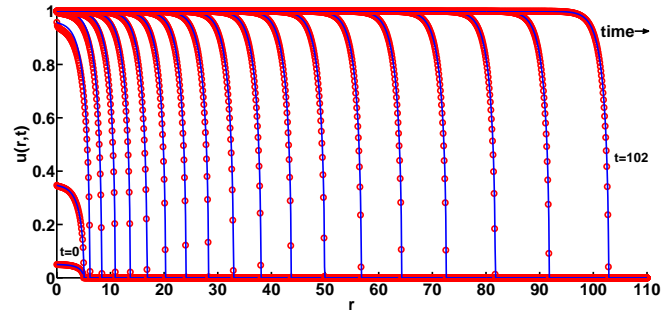}}
\caption{\label{fig:Density}
(Color online) Demonstration of evolution of the radially symmetric population density profile $u(r,t)$ \eqr{eq:main_sol_exact} in 2D ($\gamma = 1$) by comparing with the numerical solution. The solid lines represent the exact solutions and the circle markers represent the numerical solutions. The parameters are as follows: $p=2$, $\rho=0.05$ and $r_{0}=5$. The density profiles are initiated at $t=0$ and evolve until $t=102$. 
}
\end{figure}

The evolution of population density profiles, obtained from the analytical solution \eqr{eq:main_sol_exact} and the numerical solution, are demonstrated in \Fig{fig:Density}. The density initiates from a sharp profile then grows locally to the saturated value, while it spreads out to unoccupied region. At the early state, while density is small, the correction term (the last term in \Eq{eq:gen_fisher_rad3}) does not interfere the analytical density, as mentioned above. Our approximation is not accurate as the density grows to unity at early regime. However, it is seen that both of the analytical solution and the numerical solution seem to be in agreement as time and distance become large. 

\begin{figure}[h]
\centerline{\includegraphics[width=\columnwidth]{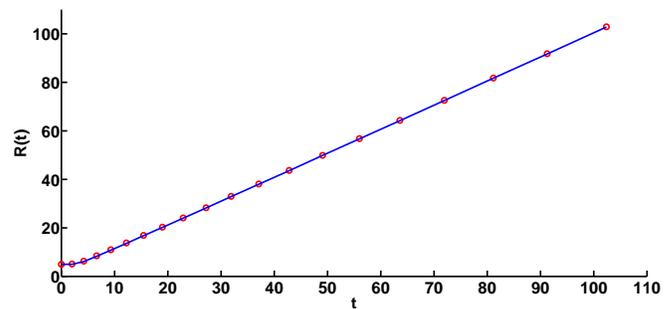}}
\caption{\label{fig:Front}
(Color online) The corresponding front position $R(t)$ extracted from the density profiles in \Fig{fig:Density}. The solid lines represent the exact solutions and the circle markers represent numerical solutions.
}
\end{figure}

The front position $R(t)$ is also measured directly from the density profiles in \Fig{fig:Density}. We notice that the small numerical deviation in density can make the front position to be shifted from the actual value. Therefore, the density that is less than $10^{-6}$ can be considered as zero in measuring the front position. The plot of analytical front position versus numerical front position is shown in \Fig{fig:Front}. By calculating $t$ for given measured $R$, the front position obtained from the density profiles satisfies \Eq{eq:front_position}. Once again, both of the analytical front position and the numerical front position are in agreement. Noting that, although it looks similar to, the data cannot be well fitted with the solution from 1D \cite{Ngamsaad2012}, for not so large distance.

In summary, we study the population dynamics that is described by the density-dependent reaction-diffusion equation, so called the extended Fisher model. We have found the approximate solution in the radial symmetric form in two- and three-dimensional space to this equation. The analytical result shows that the evolution of population density is self-similar.  At large time scale, the population density propagates as the curved traveling wave-like. The analytical solutions seem to be in agreement with the numerical solutions. Finally, it is revealed that the density profile and the propagating speed of the evolving population are influenced by the ineluctable curvature at distance that is insufficient large.



\bibliography{DDRDE_ref}

\end{document}